\documentclass[%
 reprint,
superscriptaddress,
showpacs,
 amsmath,amssymb,
 aps,
 pra,
floatfix, longbibliography ]{revtex4-1}
\usepackage{graphicx}
\usepackage{dcolumn}
\usepackage{bm}
\usepackage{soul,color}
\usepackage[colorlinks,linkcolor=blue,anchorcolor=blue,citecolor=blue,urlcolor=blue]{hyperref}
\usepackage[]{datetime}
\begin{document}

\title{{\hfill\small\rm Phys. Rev. Applied {\bf 12}, 024054 (2019)} \\%
       Designing an All-Carbon Membrane for Water Desalination}

\author{David Tom\'anek}
\email[E-mail: ]{tomanek@pa.msu.edu}
\affiliation{Physics and Astronomy Department,
             Michigan State University,
             East Lansing, Michigan 48824-2320, USA}

\author{Andrii Kyrylchuk}
\affiliation{Physics and Astronomy Department,
             Michigan State University,
             East Lansing, Michigan 48824-2320, USA}
\affiliation{Institute of Organic Chemistry,
             National Academy of Sciences of Ukraine,
             Murmanska Str. 5, 02660 Kyiv, Ukraine}

\date{\today}

\begin{abstract}
We design an all-carbon membrane for the filtration and
desalination of water. A unique layered assembly of carbon
nanostructures including graphite oxide (GO), buckypaper
consisting of carbon nanotubes, and a strong carbon fabric
provides high mechanical strength and thermal stability,
resilience to harsh chemical cleaning agents and electrical
conductivity, thus addressing major shortcomings of commercial
reverse osmosis membranes. We use {\em ab initio} density
functional theory calculations to obtain atomic-level insight into
the permeation of water molecules in-between GO layers and across
in-layer vacancy defects. Our calculations elucidate the reason
for selective rejection of solvated Na$^+$ ions in an optimized GO
membrane that is structurally stabilized in a sandwich arrangement
in-between layers of buckypaper, which are protected on both sides
by strong carbon fabric layers.
\end{abstract}



\maketitle



\section{Introduction}

Availability of potable water is one of the most pressing needs of
humankind~\cite{Elimelech2011}. Even though water is plentiful on
earth, most of it is not suited for human consumption and must be
treated beforehand~\cite{{Elimelech2011},{Werber2016}}. The key to
processing contaminated or brine water is filtration. Micro- and
nano-porous carbon has been used successfully for filtration due
to its small pore size and large surface area, combined with its
high chemical, thermal and mechanical stability as well as low
cost~\cite{{Boretti2018},{Liu2015}}. Also salty seawater can be
made potable using, among others, ultra-fine membranes. We focus
here on desalination based on reverse osmosis~\cite{Cohen2017},
which is more challenging than filtration due to the difficulty to
separate solvated ions from water on the molecular level.
State-of-the-art nano-porous membranes for water desalination
achieve selective rejection of specific ions, typically utilizing
polymers such as polysulfone. Such membranes do not operate well
at high temperatures approaching that of boiling water, pressures
of several hundred bar, and show limited resilience to chemicals
used for cleaning of bio-fouling debris~\cite{Warsinger2018}.
Carbon nanostructures have been considered in search of suitable
alternatives, but found to be rather brittle and unsuited for
membrane applications~\cite{Sun2016}. There have been calls for
atomic-level theoretical investigations~\cite{Elimelech2011} and
for a paradigm shift~\cite{{Elimelech2011},{Shahzad2019}} to
achieve serious progress in water desalination.

\begin{figure}[b]
\centering
\includegraphics[width=0.75\columnwidth]{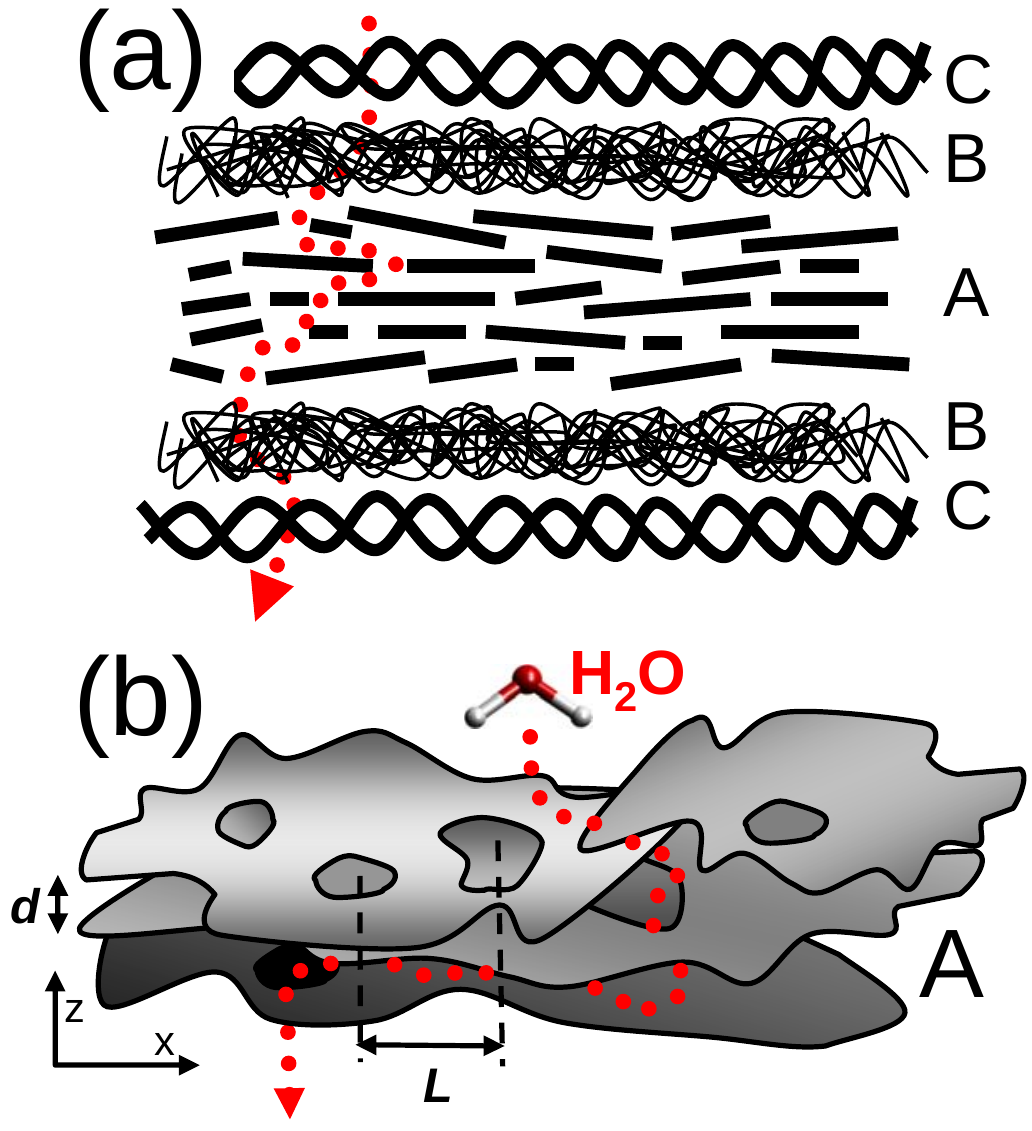}
\caption{Layered structure of the proposed all-carbon membrane
suitable for water filtration and desalination. %
(a) Schematic side view of the membrane design. Core layer {\em
A}, consisting of highly oriented graphite oxide (GO), is
sandwiched in-between layers {\em B} of buckypaper and finally
in-between layers {\em C} of strong carbon fabric. %
(b) Detailed schematic view of the defective GO layers in the core
{\em A} of the membrane, characterized by the inter-layer distance
$d$ and the typical separation $L$ between in-layer pores. %
Schematic route of an H$_2$O molecule across the membrane is
indicated by the red doted line. %
}%
\label{fig1}
\end{figure}

Here we present a design of an all-carbon membrane for the
filtration and desalination of water that should not suffer from
previously reported materials limitations. The outermost layer of
the proposed sandwich structure of the membrane would be a strong,
yet flexible fabric of carbon fibers or carbon nanotubes (CNTs),
capable of withstanding extremely high pressures used in the
reverse osmosis process. The innermost layer, which should
separate water molecules from solvated ions including Na$^+$,
would consist of layered, highly oriented graphite oxide (GO).
This core layer would be sandwiched in-between layers of
buckypaper consisting of CNTs, which would also provide cushioning
against the outermost layers. Using different polymorphs of
graphitic carbon as membrane components should provide superior
thermal and mechanical stability as well as resilience to harsh
chemicals used to clean off bio-fouling debris. To investigate the
viability of our concept, we use {\em ab initio} density
functional theory calculations to obtain atomic-level insight into
the permeation of water molecules in-between GO layers and across
in-layer vacancy defects. Our results elucidate the reason for the
reported selective rejection of solvated Na$^+$ ions in an
optimized GO membrane.


\section{Structure and functionality of the proposed all-carbon
         membrane}

A schematic view of the proposed all-carbon membrane, along with a
likely trajectory of a water molecule permeating the membrane, is
presented in Fig.~\ref{fig1}(a). In the following, we will first
discuss the properties of the outermost layers {\em B} and {\em C}
of the membrane, which are well known in the carbon community.


\subsection{Outermost layers of bucky paper and carbon fabric}

The two outermost layers {\em B} and {\em C} of the membrane in
Fig.~\ref{fig1}(a) consist of quasi-1D graphitic carbon
nanostructures including CNTs and carbon fibers. Both benefit from
extraordinary properties of graphite or graphene, including its
flexibility, in-plane tensile strength hundred times that of
steel, melting temperature close to $4,000$~K, and resilience to
harsh cleaning chemicals containing chlorine~\cite{DT227}.

The buckypaper layer {\em B} consists of entangled single-wall
CNTs with diameters of $1-2$~nm, with its structure resembling a
brillo pad~\cite{{DT227},{Rinzler98},{Sweetman2012}}. The typical
pore size, representing the inter-nanotube spacing, ranges between
$20-50$~nm. Buckypaper does not impede passage of water molecules
or ions, but blocks larger
debris~\cite{{Sweetman2012},{Rashid2014}}. The main purpose of
layer {\em B} is to contain the layered, but brittle GO with
typical flake sizes exceeding $1$~$\mu$m in the core layer {\em
A}.

The outermost carbon fabric layer {\em C} plays its most important
role in structurally supporting and containing material in layers
{\em A} and {\em B}. The pore size in layer {\em C} typically
exceeds $1$~$\mu$m and does not really matter as long as it
contains the buckypaper layer. Since the pore sizes exceed those
of buckypaper, this layer also does not impede passage of water
molecules or ions. The strength and flexibility of the carbon
fabric, which also protects the membrane from stone impact, is
provided by CNTs, which can be harvested, spun into
yarns~\cite{Zhang04} and woven to a fabric~\cite{Luo19}. Presence
of CNTs has been shown to reduce bio-fouling of
membranes~\cite{Rashid2017}. Since layers {\em B} and {\em C} are
electrically conductive, applying a voltage may furthermore assist
in rejecting ions~\cite{{Ho2018},{Perez-Roa2009},{Park2006}}.


\subsection{Layered GO}

Key to the selective rejection of ions is the core layer {\em A}
of the membrane consisting of GO. The structure of GO consists of
graphene layers that have been chemically functionalized by
epoxy-O and OH groups~\cite{Boehm1961}. A simplified schematic
image of a realistic GO structure is presented in
Fig.~\ref{fig1}(b). Unlike hydrophobic graphite and graphene, GO
is hydrophilic and requires water for stabilization. In comparison
to graphite, GO also has a lower bio-fouling
tendency~\cite{Hu2016}.

The two commonly used approaches to synthesize GO are the
Hummer~\cite{Hummers1958} and the Brodie~\cite{Brodie1859}
process. Both techniques yield an ordered material containing a
substantial fraction of in-layer pores, seen in
Fig.~\ref{fig1}(b). For a long time, GO has been considered a
viable material to form membranes, especially suitable for water
desalination based on reverse osmosis~\cite{Boehm1961}. GO has
been found very useful for the filtration and purification of
water~\cite{Akhavan10} and other substances, which have to cross
many pores in many layers. GO membranes have come back into the
focus of research due to progress in this direction. Due to rather
small grain size, as-produced GO is typically regarded a powder.
Related multilayer graphene oxide (MLGO), a new term coined for
layered GO with large grains, has indeed proven useful for water
purification and desalination~\cite{Ventrice2009}. A significant
improvement of the nanofiltration capability of GO has been
achieved by shear alignment~\cite{Akbari16}. We believe that
homogenization of the GO sample by ball milling followed by shear
alignment may optimize the performance of GO for water
desalination.

Obviously, controlling the microscopic structure and layer
alignment in GO and understanding the flow mechanism of water
molecules in-between layers of graphite~\cite{Cicero2008} and
GO~\cite{{Korobov16},{Gogoi18}} are crucial for further progress.
Molecular dynamics simulations with model
potentials~\cite{{Lerf06},{Yang16}} and neutron diffractions
studies~\cite{{Buchsteiner2006},{Lerf06}} have provided initial
insight into the system. Still, atomic-scale information about the
mechanism of water permeation and selective ion rejection by GO
membranes is either absent or extremely limited.


\subsection{Known facts about the permeation of H$_2$O molecules
            and salt ions in GO}

Presence of in-layer pores in the layered structure of GO is a
necessary prerequisite for water permeation since no atoms could
penetrate through the atomic lattice of an ideal membrane
consisting of infinitely large, defect-free layers. We distinguish
in-layer pores, consisting of extended vacancies and defects in a
layered structure, from slit pores associated with the interlayer
region that can accommodate external atoms and molecules. To allow
water to permeate, pore sizes in the membrane must exceed the van
der Waals diameter $d_{vdW}$(H$_2$O)$=2.8$~{\AA} of a H$_2$O
molecule~\cite{Jorgensen83}. The key challenge is to allow only
water and not salt ions to pass through the membrane by limiting
the size and chemical termination of in-layer pores.

Description of ion transport across in-layer pores in GO must
necessarily consider the fact that ions are hydrated and that the
ionic hydration shell may change during the passage. It would
therefore be misleading to compare the observed bare
radii of salt ions~\cite{Volkov1997} %
$r{\rm{(Na}}^+{\rm{)}}=1.17$~{\AA}, %
$r{\rm{(K}}^+{\rm{)}}=1.49$~{\AA}, and %
$r{\rm{(Cl}}^-{\rm{)}}=1.64$~{\AA} %
to the van der Waals radius
$r$(H$_2$O)$=1.4$~{\AA} %
of a water molecule~\cite{Jorgensen83}.

In aqueous environment, all ions are surrounded by water
molecules, which form a tightly bound hydration shell. Neutron
diffraction studies~\cite{Mancinelli2007} and DFT
calculations~\cite{Bankura2013} agree on the hydration shell radii
that are slightly larger than the ion-oxygen distances %
$d$(Na$^{+}-$O)$=2.3-2.4$~{\AA}, %
$d$(K$^{+}-$O)$=2.7-2.8$~{\AA}, and %
$d$(Cl$^{-}-$O)$=3.2$~{\AA}. %
Unlike the cations, hydrated Cl$^-$ ions form hydrogen-bonded
bridges with water molecules~\cite{Mancinelli2007}. %
The hydration shell of Na$^+$ and K$^+$ cations contains about $6$
water molecules, that of larger Cl$^-$ anions about $7$ water
molecules. The values of the sequential detachment energies of
water molecules from the hydration shells of many ions are now
well established~\cite{NIST-hydration}. For ions of interest here,
the detachment of all water molecules requires %
$E_{hyd}$(Na$^{+}$)$=4.03$~eV, %
$E_{hyd}$(K$^{+}$)$ =3.46$~eV, and %
$E_{hyd}$(Cl$^{-}$)$=3.27$~eV. %
These hydration energy values are significantly higher than the
hydration energy of a water molecule~\cite{Pang2014} %
$E_{hyd}{\rm{(H}}_2{\rm{O)}}=0.43$~eV and have to be considered
seriously in any desalination study.

Steric and energetic considerations suggest that with their
largest hydration shell, Cl$^-$ anions should pass with much less
ease than the Na$^{+}$ and K$^{+}$ cations across a membrane
containing nanopores. This expectation has been confirmed by the
experimental finding~\cite{Rollings2016} that K$^{+}$ and Na$^{+}$
behave in a similar way, and that K$^{+}$ ions pass 100 times
faster than Cl$^{-}$ ions through a membrane. With their even
larger hydration shells, divalent cations Ca$^{++}$ and Mg$^{++}$
have been observed to pass still more slowly than monovalent
cations~\cite{Rollings2016}. Therefore, we will focus on the
selective passage of water molecules and monovalent cations in
this study.

Passage of ions through a membrane may be suppressed by limiting
the pore size to that of the ionic hydration shell, with $6$~{\AA}
for hydrated Na$^+$ representing a minimum
value~\cite{{Mahler2012},{Bankura2013}}. The major challenge in
the desalination field is production of membranes with a very
narrow range of pore sizes between $0.3-0.8$~nm that enable
passage of water, but not of hydrated ions.

In this study, we combine reported observations with original
atomic-scale calculation results that should elucidate the
permeation of water and specific rejection of ions in a GO
membrane, which have eluded a consistent microscopic description
so far.


\subsection{Cleaning of bio-fouled all-carbon membrane}

Bio-fouling of membranes can never be prevented and poses a
significant problem in the RO
process~\cite{{Warsinger2018},{Rachman2013},{Zhang2013},{Hu2016}}.
Conventional cleaning agents, which contain chlorine, not only
remove bio-fouling agents, but also attack currently used polymer
membranes, limiting their useful lifetime. One major benefit of an
all-carbon membrane is the known resilience of carbon materials to
cleaning agents containing chlorine and to other chemicals.

Stability of carbon membrane materials under temperatures well
beyond $1,000^\circ$C suggests an unconventional alternative to
chemical cleaning. For the purpose of cleaning, an all-carbon
membrane could be separated from its holder, suspended in an
oxygen-free nitrogen atmosphere, and heated resistively to high
temperatures, when all bio-fouling deposits would decompose. The
most stable product of the decomposition process should be
non-specific carbon nanostructures, which should provide
additional structural reinforcement to the membrane material and
potentially improve its performance without clogging it.


\section{Computational approach}

Our computational approach to study liquid water and Na$^+$ ions
interacting with chemically functionalized, defective graphite and
graphite oxide is based on \textit{ab initio} density functional
theory (DFT) as implemented in the {\textsc{SIESTA}}~\cite{SIESTA}
and {\textsc{VASP}}~\cite{VASP,VASPPAW} codes. %
While computationally rather demanding, DFT calculations are free
of adjustable parameters and provide bias-free predictions for
molecular and solid systems. In this, our approach differs from
parameterized force fields that are computationally faster, but
offer limited transferability among different systems. %
We described the effect of electron exchange and correlation using
the nonlocal Perdew-Burke-Ernzerhof (PBE)~\cite{PBE} functional
and compared selected results to those based on the Local Density
Approximation (LDA)~\cite{{Ceperley1980},{Perdew81}}. The DFT-PBE
total energy functional has been used extensively to provide an
unbiased description of water and its interaction with
solids~\cite{{Cicero2008},{Ambrosetti2011}}. We have used periodic
boundary conditions throughout the study. Our {\textsc{SIESTA}}
calculations used norm-conserving Troullier-Martins
pseudopotentials~\cite{Troullier91}, a double-$\zeta$ basis
including polarization orbitals, and a mesh cutoff energy of
$180$~Ry to determine the self-consistent charge density, which
provided us with a precision in total energy of
${\lesssim}2$~meV/atom. The {\textsc VASP} calculations were
performed using the projector augmented wave (PAW)
method~\cite{VASPPAW} and $400$~eV as energy cutoff. The
reciprocal space has been sampled by a uniform $k$-point
grid~\cite{Monkhorst-Pack76}, as noted specifically. Systems with
very large unit cells have been represented by the $\Gamma$ point
only. Geometries have been optimized using the conjugate gradient
(CG) method~\cite{Hestenes1952}, until none of the residual
Hellmann-Feynman forces exceeded $10^{-2}$~eV/{\AA}. Equilibrium
structures and energies based on {\textsc{SIESTA}} have been
checked against values based on the {\textsc{VASP}} code.
Microcanonical (NVE) and canonical (NVT) molecular dynamics (MD)
calculations were performed using short $0.3$~fs time steps, which
were sufficiently short guarantee energy conservation in
microcanonical ensembles.

To represent the effect of pressure on the permeation of water and
Na$^+$ ions, we modified the {\textsc{SIESTA}} source code to
subject selected atoms to a constant force. To preserve
interatomic distances and obtain a constant acceleration within a
group of selected atoms, we subject each atom in the group to the
force ${\bf{F}}={\bf{F_0}}{\times}m_a$, where $m_a$ is the atomic
mass in a.m.u. These driving forces, which typically amount to
$|{\bf{F_0}}|{\lesssim}2.5{\times}10^{-2}$~eV/{\AA}, %
induce water flow along and across defective graphene and graphite
oxide layers in our MD simulations.


\section{Results}

All our results reported in the following are based on {\em ab
initio} total energy calculations and corresponding MD
simulations. We use DFT to obtain static equilibrium structures
including their energies and DFT-based MD simulations to describe
the evolution of a system in time. Such numerical simulations are
often referred to as computational experiments as they produce
vast quantities of data, which need to be further analyzed. Both
in real and computational experiments, error bars represent the
uncertainty in the reported values. In computer simulations, error
bars may be obtained from ensemble averages or estimated based on
finite-size fluctuations in finite-time MD simulations.


\subsection{Properties of liquid water}

Whereas isolated H$_2$O molecules are well understood, correct
description of liquid water has remained a challenging problem for
decades~\cite{Chen2017}. Very many energy functionals have been
developed to describe the delicate interplay among the strong and
weak forces that determine the behavior of liquid water, but none
has been able to satisfactorily reproduce all aspects of its
behavior and interaction with solids in an unbiased manner. In
spite of their relatively high requirement on computational
resources, we use here {\em ab initio} DFT calculations to provide
an unbiased description of water interacting with carbon-based
membranes.

Our calculated hydration energy of a water molecule
$E_{hyd,theo}{\rm{(H}}_2{\rm{O)}}=0.408$~eV %
agrees well with the observed value~\cite{Pang2014}
$E_{hyd,expt}{\rm{(H}}_2{\rm{O)}}=0.430$~eV. %
The calculated gravimetric density of water
$\rho_{theo}{\rm{(H}}_2{\rm{O)}}=0.9$~g/cm$^3$
agrees with previously reported DFT-PBE
values~\cite{{Miceli2015},{Gaiduk2015},{Chen2017}} of %
$0.81-0.87$~g/cm$^3$, but underestimates
the observed value at room temperature~\cite{Franks2000} %
$\rho_{expt}{\rm{(H}}_2{\rm{O)}}=1.0$~g/cm$^3$.

To obtain insight into the dynamics of liquid water, we first
considered its viscous behavior. To do so, we used the
Einstein-Smoluchowsky equation, which relates the self-diffusion
coefficient $D$ of a Brownian particle in a
${\tilde{d}}$-dimensional space to its trajectory ${\bf r}(t)$ by
\begin{equation}
<|{\bf r}(t)-{\bf r}(0)|^2> = 2 {\tilde d} D t \;. %
\label{eq1}
\end{equation}
The dimensionality of bulk water in this equation is
${\tilde{d}}=3$. We performed a limited-size MD simulation of 30
H$_2$O molecules per unit cell and traced the positions of water
molecules by their center of mass as a function of time. Our
simulation results shown in Fig.~\ref{fig3}(e) suggest
$D_{theo}=(1.0{\pm}0.8){\times}10^{-5}$~cm$^2$/s for the
self-diffusion coefficient. In spite of the large uncertainty
caused by limited simulation time and the significant finite-size
fluctuations, the estimated value of $D$ is somewhat lower than
the observed value~\cite{Holz2000}
$D_{expt}=2.3{\times}10^{-5}$~cm$^2$/s$=2.3{\times}10^{-9}$~m$^2$/s %
in bulk liquid H$_2$O at $25^\circ$C. The fact that DFT-PBE MD
simulations typically underestimate the value of $D$, with results
ranging from $0.1{\times}10^{-5}-0.6{\times}10^{-5}$~cm$^2$/s, has
been reported previously~\cite{{Chen2017},{Gillan2016b}}. Also,
the higher initial slope of the mean square displacement in our MD
simulation, caused by initial ballistic motion of H$_2$O, agrees
with previously reported behavior~\cite{Youssef2011}.


\subsection{Representation of graphite oxide (GO)}

\begin{figure}[h]
\centering
\includegraphics[width=1.0\columnwidth]{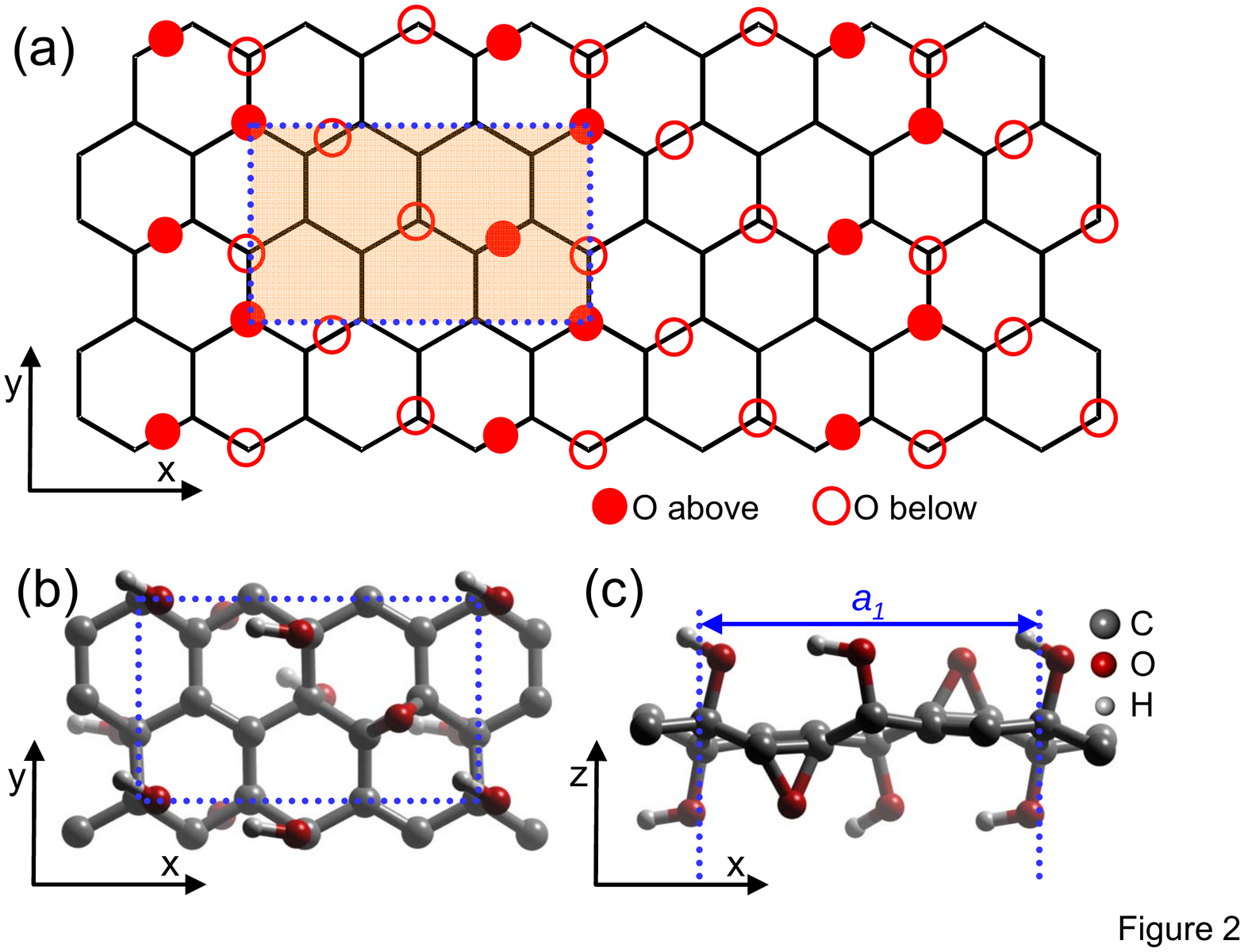}
\caption{%
Monolayer of graphite oxide in the Lerf-Klinowski
model~\cite{Lerf1998} according to Ref.~\cite{Buchsteiner2006}.
(a) Schematic representation in top view, indicating the position
of epoxy-O in bridge sites and OH-groups in on-top sites above and
below the graphene monolayer. %
Ball-and-stick models of optimized GO in %
(b) top view and %
(c) side view, %
indicating a substantial degree of wrinkling. The primitive unit
cell is delimited by the dotted blue line and indicated by
shading in (a).
} %
\label{fig2}
\end{figure}

Whereas the honeycomb lattice of graphene, with atoms separated by
$1.42$~{\AA}, has been known for a long time~\cite{Graphite1916},
its oxidized counterpart -- GO -- is not as clearly defined and
contains covalently attached epoxy-O and OH-groups that may vary
in their spatial distribution. GO may contain isolated
non-oxidized graphitic regions amounting to less than $15\%$ by
area~\cite{Erickson10}. The most commonly used representation of
fully oxidized GO is the Lerf-Klinowski
model~\cite{{Lerf1998},{Buchsteiner2006}} shown in
Fig.~\ref{fig2}.


\subsection{Interaction of H$_2$O molecules with graphite and GO layers}

\begin{figure*}[t]
\centering
\includegraphics[width=1.5\columnwidth]{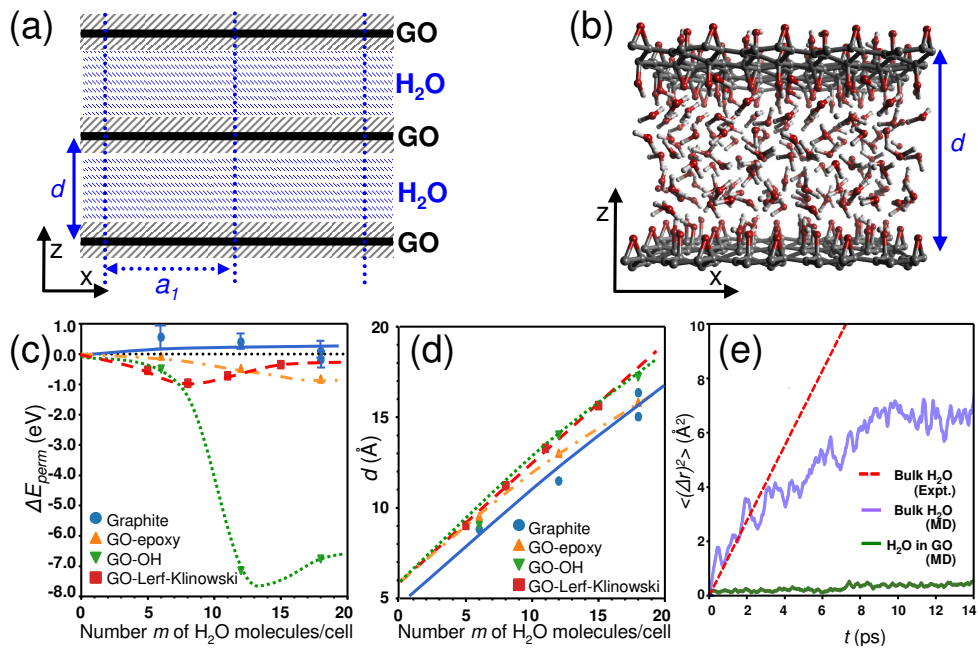}
\caption{%
Permeation of water in GO. %
(a) Schematic side view of the periodic structure of GO layers
separated by water. The GO unit cell of width $a_1$ and height $d$
is the same as in Fig.~\protect\ref{fig2}. %
(b) Snap shot of the arrangement of H$_2$O molecules in a
$14.72$~{\AA} wide segment of GO with the interlayer separation
$d=12.85$~{\AA}. The dashed lines indicate hydrogen bonds. %
(c) Energy gain ${\Delta}E$ associated with the presence of $m$
water molecules in the interlayer region. %
(d) Optimized height $d$ of unit cells containing $m$ water
molecules. %
(e) Comparison of the mean-square displacement of H$_2$O molecules
in bulk water and in the constrained space in-between GO layers
shown in panel (b). Extrapolation based on experimental data in
bulk water is shown for comparison by the red dashed line.
}%
\label{fig3}
\end{figure*}

As a water molecule adsorbs on graphene or graphite, it gains the
adsorption energy $E_{ad,theo}{\rm{(H}}_2{\rm{O)/MLG}}=0.14$~eV
according to our DFT-PBE results. This is similar to the observed
value~\cite{Avgul1970}
$E_{ad,expt}{\rm{(H}}_2{\rm{O)/MLG}}{\lesssim}0.20$~eV, %
which is likely overestimated~\cite{Ambrosetti2011}. The
interaction energy of a 2D water monolayer with graphene,
amounting to $0.02$~eV per H$_2$O molecule, %
is one order of magnitude smaller. Adsorption of initially free
water molecules and their subsequent condensation to a 2D
monolayer on graphene provides a larger energy gain of $0.39$~eV
per water molecule.
All these values are significantly lower than the hydration energy
$E_{hyd,theo}{\rm{(H}}_2{\rm{O)}}=0.408$~eV of a water molecule in
water. Energetic preference of H$_2$O to be surrounded by other
water molecules rather than graphitic carbon makes graphene and
graphite hydrophobic.

We could not confirm the reported formation of 2D square ice on a
graphene monolayer\cite{Algara-Siller2015},
which have been questioned subsequently~\cite{Zhou15},
based on our total energy results, which suggest energy
differences of only few meV/molecule due to molecular ordering.
Such small energy differences are consistent with previously
reported theoretical results~\cite{Boukhvalov2013} indicating easy
sliding of H$_2$O molecules in-between graphene layers.

We find the adsorption energy of a water molecule on GO to be
significantly higher than on graphene and graphite. Depending on
the adsorption site, typical values range near
$E_{ad,theo}{\rm{(H}}_2{\rm{O)/GO}}=0.73$~eV, %
exceeding the hydration energy of H$_2$O significantly. This
indicates that water molecules prefer energetically to be near GO
rather than surrounded by water only, making GO hydrophilic on the
molecular level. Our findings for isolated water molecules are
only indicators that should not be taken as substitutes for the
behavior of liquid water.

In a more realistic description, we considered the permeation of
water in a carbon-based layered system, namely GO or graphite,
using the periodic geometry shown in Fig.~\ref{fig3}(a). A snap
shot of water molecules arranged in-between GO layers is shown in
Fig.~\ref{fig3}(b). Per unit cell with in-layer area $A$ and
height $d$, the water permeation energy ${\Delta}E_{perm}$
associated with the transfer of $m$ water molecules from liquid
water to the interlayer region can be estimated by obtaining the
energy difference between the water-permeated system, bulk water
and the bulk dry carbon-based system. The carbon-based system will
be hydrophobic if ${\Delta}E_{perm}>0$ and hydrophilic if
${\Delta}E_{perm}<0$, indicating that water molecules prefer
energetically the interlayer space in the carbon-based system to
bulk water. Beyond a critical content of water, amounting to many
H$_2$O molecules per unit cell and corresponding to several layers
of water in the inter-layer space, ${\Delta}E_{perm}$ should
approach a constant value describing non-interacting GO layers
solved in bulk water.

\begin{figure}[b]
\centering
\includegraphics[width=1.0\columnwidth]{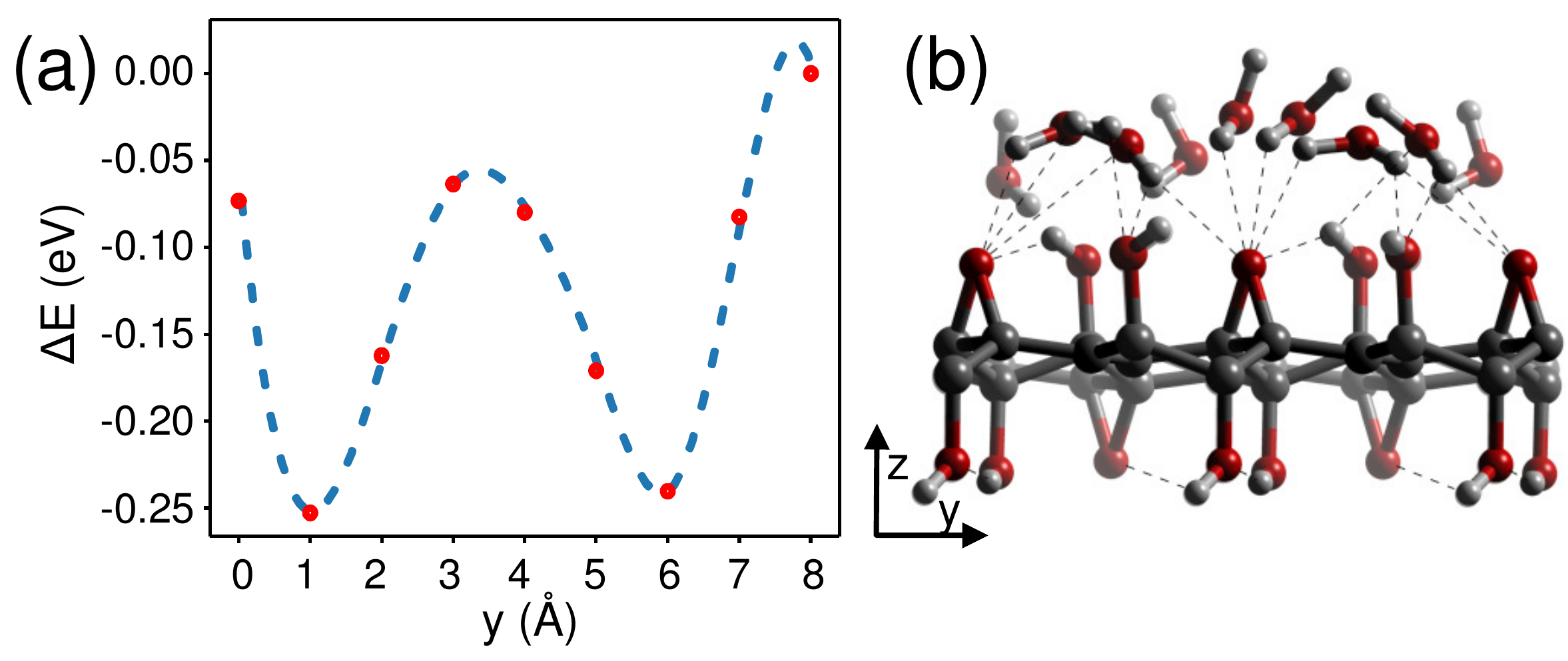}
\caption{%
Diffusion of a water molecule confined in the $x-z$ plane normal
to GO layers along the GO layers within a slit pore in-between
the layers. %
(a) Energy change ${\Delta}E$ as the water molecule diffuses along
the armchair direction $y$ of GO. %
(b) Snap shots of water molecule configurations along the
    diffusion path, corresponding to the
    simulation
    data points in (a),
    with hydrogen bonds indicated by the dashed lines.
The line connecting the simulation
data points in (a) is a guide to the eye. %
} %
\label{fig4}
\end{figure}

\begin{figure*}[t]
\centering
\includegraphics[width=2.0\columnwidth]{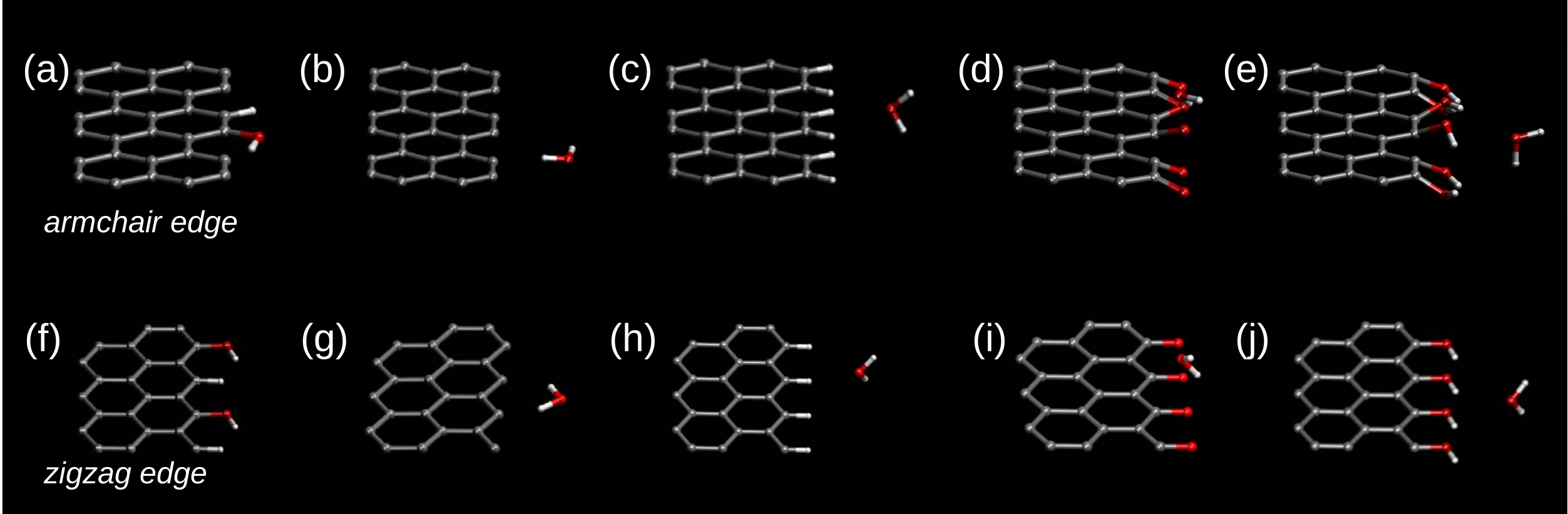}
\caption{%
Optimum geometry of an H$_2$O molecule interacting with armchair
(a-e) and zigzag (f-i) edges of a graphene monolayer. Bare edges
support both dissociative chemisorption (a,f) and physisorption
(b,g). Only physisorption occurs on H-terminated (c,h),
O-terminated (d,i) and OH-terminated (e,j) edges. %
} %
\label{fig5}
\end{figure*}

\begin{figure*}[t]
\centering
\includegraphics[width=2.0\columnwidth]{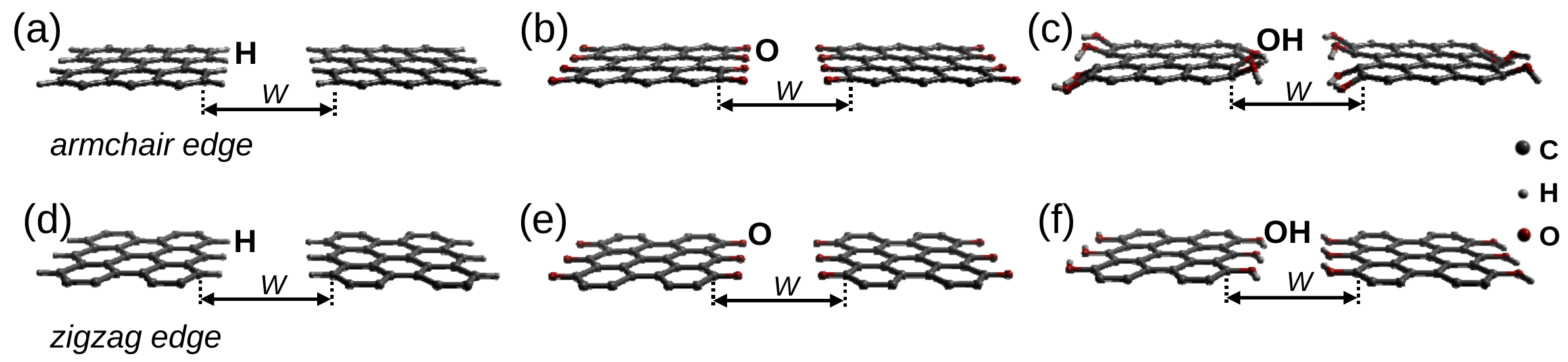}
\caption{%
Optimum geometry of an idealized in-layer pore formed by cleaving
a graphene monolayer along the armchair or the zigzag direction.
The armchair (a-c) and zigzag (d-f) edges may be chemically
terminated by -H (a,d), -O (b,e), or -OH (c,f) groups. The pore
width $W$ is defined by the separation of closest carbon atoms at
opposing pore edges.
} %
\label{fig6}
\end{figure*}

We found the energy balance in the calculation of
${\Delta}E_{perm}$ to depend delicately on the arrangement of
water molecules in the interlayer region. To compensate for this
uncertainty to some degree, we first calculated
${\Delta}\tilde{E}_{perm}$ not with respect to bulk H$_2$O, but
rather free-standing layers of H$_2$O in exactly the same
configuration as in the interlayer space of the carbon-based
layered system. Then, we determined
${\Delta}E_{perm}={\Delta}\tilde{E}_{perm}+2{\gamma}A$, where $A$
is the area of the unit cell in the plane of the layers and
$\gamma$ is the surface tension of liquid water, which in turn
depends on its surface structure in calculations. We used
$A=41.9$~{\AA$^2$} for optimized graphene with a 16-atom
supercell, $A=43.1$~{\AA$^2$} for GO terminated with O-epoxy
groups, $A=41.0$~{\AA$^2$} for GO terminated with OH-groups, and
$A=32.3$~{\AA$^2$} for the Lerf-Klinowski
model~\cite{{Lerf1998},{Buchsteiner2006}} of GO shown in
Fig.~\ref{fig2}. Based on the density distribution of water
molecules and structural snap shots such as Fig.~\ref{fig3}(b), we
find that distinguishable water layers contain typically 5
molecules per unit cell in our calculation, which helps in
correlating the number $m$ of water molecules with the number of
water layers.

Theoretical estimates of $\gamma$ range from
$6.6{\times}10^{-2}$~J/m$^2$ based on DFT-PBE MD
simulations~\cite{Ohto2019} to the estimated value
$\gamma{\lesssim}65.9{\times}10^{-2}$~J/m$^2$
based on our DFT-PBE structure optimization studies for different
geometries. The observed value~\cite{H2O-CRC86}
$\gamma=7.3{\times}10^{-2}$~J/m$^2$ at $20^\circ$C
falls into this value range and so does the value
$\gamma=14.0{\times}10^{-2}$~J/m$^2$ we selected to estimate the
water permeation energy.

Our results for ${\Delta}E_{perm}$ as a function of the number $m$
of the water molecules per unit cell are presented in
Fig.~\ref{fig3}(c). Based on the sign of ${\Delta}E_{perm}$, we
conclude that the H$_2$O/GO system is hydrophilic and the
H$_2$O/graphite system is hydrophobic in agreement with
observation. The simulation data points are subject to variations
caused by the amorphous structure of liquid water that is
continuously changing. Among the GO systems, we find GO-OH to be
most and GO-epoxy to be least hydrophilic, with the Lerf-Klinowski
model containing both OH- and O-epoxy groups to lie in-between.
There even appears to be preference for about two water layers in
GO according to the Lerf-Klinowski model, in agreement with
observation~\cite{Korobov16}, and about three water layers in
GO-OH, probably due to favorable registry of water molecules with
the functionalized carbon layers. We find that energy changes in
the system are largest during the initial permeation of water
in-between the layers, eventually reaching a saturation value.
With $m{\approx}5$ water molecules per water layer in the unit
cell, we may consider systems with $m{\gtrsim}15$ to represent a
fully hydrated system.

We present the optimized interlayer distance $d$ as a function of
water filling in Fig.~\ref{fig3}(d). The increase of $d$ due to
water permeation, starting from $d{\approx}6$~{\AA} in dry GO, is
called swelling. In an RO system containing a GO-based membrane,
swelling will be limited by the pressure difference between the
two sides of the membrane and by inter-layer bonds that hold a
defective GO membrane together.

In agreement with previously reported
theoretical~\cite{Bankura2013} and experimental
data~\cite{Mahler2012}, we find the optimum size of the Na$^+$
hydration shell to be close to $6$~{\AA}. Since the interlayer
distance $d$ in GO permeated with water is significantly larger,
hydrated Na$^+$ ions are likely to permeate in and propagate
within the interlayer region in a similar way as bulk water.

To learn more about water flow within slit pores in GO, we first
studied the diffusion of a water molecule along the armchair
direction in the GO layer, aligned with the $y-$axis in
Fig.~\ref{fig2}(a). To obtain bias-free energy information along
the diffusion path, we performed a set of global optimizations
while constraining the $y-$coordinate of the O atom of the water
molecule and allowing this atom to move only within the $x-z$
plane normal to GO layers. The energy change ${\Delta}E$ along the
trajectory is shown in Fig.~\ref{fig4}(a) and a superposition of
corresponding snapshots of the molecule is shown in
Fig.~\ref{fig4}(b). Due the activation barriers for water
propagation along the slit pore, the self-diffusion coefficient is
reduced significantly with respect to the bulk water value, as
seen in Fig.~\ref{fig3}(e). Our estimated value of the
self-diffusion coefficient of water in GO is
$D_{theo}=8{\times}10^{-7}$~cm$^2$/s. The significant reduction in
the calculated value of $D$ from bulk water to water in GO is
consistent with the observation~\cite{{Raviv2001a},{Major2006a}}
that $D$ is several times lower for water under both hydrophilic
and hydrophobic confinement than in bulk water.

\begin{figure*}[t]
\centering
\includegraphics[width=1.6\columnwidth]{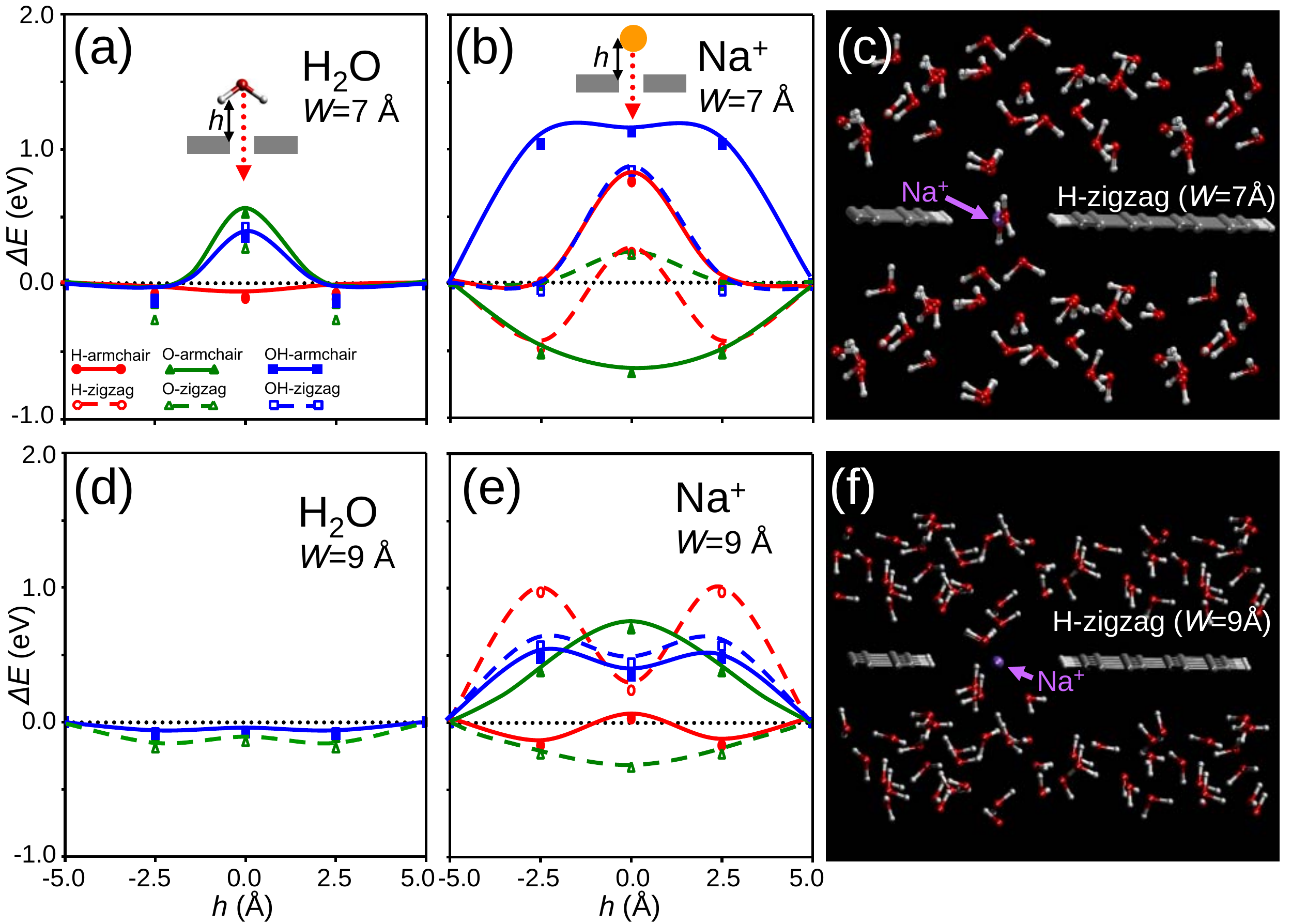}
\caption{%
Permeation of an isolated H$_2$O molecule and of hydrated Na$^+$
ions through in-layer pores with different pore widths $W$ and
different chemical terminations. $h$ represents the vertical
distance between the centroid of H$_2$O or the ion and the closest
carbon layer. Lines connecting the simulation data points are
guides to the eye.
} %
\label{fig7}
\end{figure*}

\begin{table}[b!]
\caption{%
Adsorption energy $E_{ad}$ of an isolated H$_2$O molecule at
graphene edges. Reported results are for bare as well as H-, O-
and OH-terminated armchair and zigzag edges. %
}%
\begin{tabular}{lcc} %
\hline \hline
   \textrm{Edge termination} %
 & \textrm{$E_{ad}$(armchair)} %
 & \textrm{$E_{ad}$(zigzag)} %
 \\
\hline%
  {Bare (H$_2$O$\rightarrow$H$_{ad}$+OH)} %
& {$-4.75$~eV} 
& {$-5.27$~eV} 
\\
  {Bare} %
& {$-0.02$~eV} 
& {$-0.21$~eV} 
\\
  {H-terminated} %
& {$-0.10$~eV} 
& {$-0.03$~eV} 
\\
  {O-terminated} %
& {$-0.14$~eV} 
& {$-0.10$~eV} 
\\
  {OH-terminated} %
& {$-0.08$~eV} 
& {$-0.03$~eV} 
\\
\hline \hline %
\end{tabular}
\label{table1}
\end{table}

We should note that self-diffusion under ambient conditions is
unrelated to the pressure-driven propagation of water molecules
along the slit pores. Still, the related reduction of the mobility
of water molecules adjacent to the GO layers is another validation
of the assumptions used in the Hagen-Poisseuille description of
pressure-driven flow of liquids through pores and
pipes~\cite{Hunt2014}.

To study pressure-induced water permeation along slit pores, we
filled the space in-between GO layers, defined in
Fig.~\ref{fig3}(a), partly with water. To observe the formation of
a meniscus and the dynamics of the propagating water front, we
separated segments completely filled with water by water-free
segments. All atoms in the water molecules were subject to the
force $F=F_0{\times}m_a$, where $m_a$ is the atomic mass and
$F_0=6.25{\times}10^{-3}$~eV/{\AA}, acting along the slit pores.
We performed MD simulations of the evolution of the system lasting
about $1$~ps in total and followed the acceleration of the water
molecules. Unlike in unconstrained bulk water, water molecules
in-between GO layers were slowed down by a drag force
$F_{0,drag}$, acting on both H and O atoms and originating in the
interaction between the moving molecules and the GO layers. Movies
of the propagation of water molecules along the slit pores are
presented in Section A of the Supporting Information
(SI)~\cite{desal19-SI}.

For one monolayer of water, the distance between GO layers was
fixed at $d=9.03$~{\AA} and the drag force was found to be
$F_{0,drag}=5.0{\times}10^{-3}$~eV/{\AA}, almost $80\%$ of the
driving force. For three monolayers of water in-between GO layers
separated by $d=12.9$~{\AA}, the drag force dropped to
$F_{0,drag}=2.5{\times}10^{-3}$~eV/{\AA}, about half the value for
a water monolayer. For the sake of comparison, we also considered
a monolayer of water contained in-between graphene layers
separated by $d=6.57$~{\AA}. Applying the same accelerating force
as in GO, we found a drag force of only
$F_{0,drag}=0.3{\times}10^{-3}$~eV/{\AA}, one order of magnitude
smaller than for water in GO.

We interpret these results in the following way. Whereas the water
layers next to GO are nearly immobile, other water molecules in
the middle of the slit pore are not affected much by the
constraints. This is different in (hydrophobic) graphite, where
none of the water layers would be slowed down by contact with
graphene.


\subsection{Interaction of H$_2$O molecules with defects
            in graphite and GO}

As mentioned above, presence of defects and pores in graphite and
GO is essential for the permeation of water through corresponding
membranes. The two characteristic edges of extended in-layer pores
of interest here are the armchair and zigzag edge, obtained by
cleaving graphene along the corresponding direction. Under ambient
conditions, graphitic edges are not bare, but chemically
terminated. The interaction of water molecules with these edges
depends to a significant degree on the termination with hydrogen,
hydroxy, and oxygen groups. Presence of these groups in the edge
termination of GO has been established~\cite{He1998} and their
functionality should be similar to that in a graphene edge. The
optimum geometries of water molecules interacting with graphitic
edges is depicted in Fig.~\ref{fig5} and our numerical results for
the adsorption energies of water molecules are summarized in
Table~\ref{table1}.

Energetically most favorable for the water molecule near a bare
graphene edge is dissociative adsorption shown in
Figs.~\ref{fig5}(a) and \ref{fig5}(f), which splits H$_2$O and
binds the resulting OH-radical and hydrogen to the edge covalently
with a net energy gain of $4-6$~eV according to
Table~\ref{table1}. The metastable physisorbed structures, shown
in Fig.~\ref{fig5}(b) and \ref{fig5}(g), are bound much more
weakly according to Table~\ref{table1}, similar to the adsorption
on H-, O- and OH-terminated edges. Among these, there is a small
energetic preference for the functionalized armchair edges.


\subsection{Representation of in-layer pores in graphite and GO}

In our numerical study, we represent in-layer pores in GO by an
infinite array of graphene nanoribbons with chemically
functionalized armchair and zigzag edges and changing the in-layer
separation width $W$. The structure of such pores is shown in
Fig.~\ref{fig6}.


\subsection{Permeation of H$_2$O and ions through
            in-layer pores in GO}

The key functionality of GO membranes in the desalination process
is in allowing water molecules to pass through in-layer pores,
while suppressing passage of hydrated salt ions such as Na$^+$.
Since the interlayer distance in GO permeated with water is
significantly larger than the size of hydrated Na$^+$ ions, we
must consider the presence of both Na$^+$ ions and H$_2$O
molecules in the interlayer region in GO. Then, the passage of
hydrated Na$^+$ ions across in-layer pores needs to be
investigated in the same fashion as that of water molecules.

Most important for the passage of molecules and ions through an
in-layer pore is its geometry, including the edge termination. As
mentioned before, the molecular sieving process is not affected by
the layer geometry beyond the atomically close vicinity of the
pore. Thus, we represent in-layer pores in defective GO layers by
infinite arrays of graphene nanoribbons with armchair and zigzag
edges, shown in Fig.~\ref{fig6}, which are better defined and
easier to treat computationally. We considered pores of different
width $W$, defined by the separation of closest carbon atoms at
pore edges, terminated with hydrogen, hydroxy, and oxygen
groups~\cite{He1998}.

Our results for the passage of isolated water molecules and
hydrated Na$^+$ ions through different in-layer pores are
presented in Fig.~\ref{fig7}. Corresponding movies of permeating
water molecules are presented in Section B of the SI and those of
permeating hydrated ions in Sections D and E of the
SI~\cite{desal19-SI}. To provide a realistic description of the
activated process, we performed global optimization of the system
while only fixing in-layer carbon atoms and the height $h$ of the
molecule or ion above the in-layer pore. As seen in
Figs.~\ref{fig7}(a) and \ref{fig7}(b), passage of both H$_2$O and
Na$^+$ through most narrow pores with $W=7$~{\AA} is energetically
activated, independent of edge geometry and termination.
Activation-free passage is only possible through $W=7$~{\AA} pores
with H-terminated armchair edges for H$_2$O and O-terminated
armchair edges for Na$^+$. While passing an in-layer pore, water
molecules lie preferentially in a plane normal to the carbon layer
that passes through the pore.

\begin{figure*}[t]
\centering
\includegraphics[width=1.5\columnwidth]{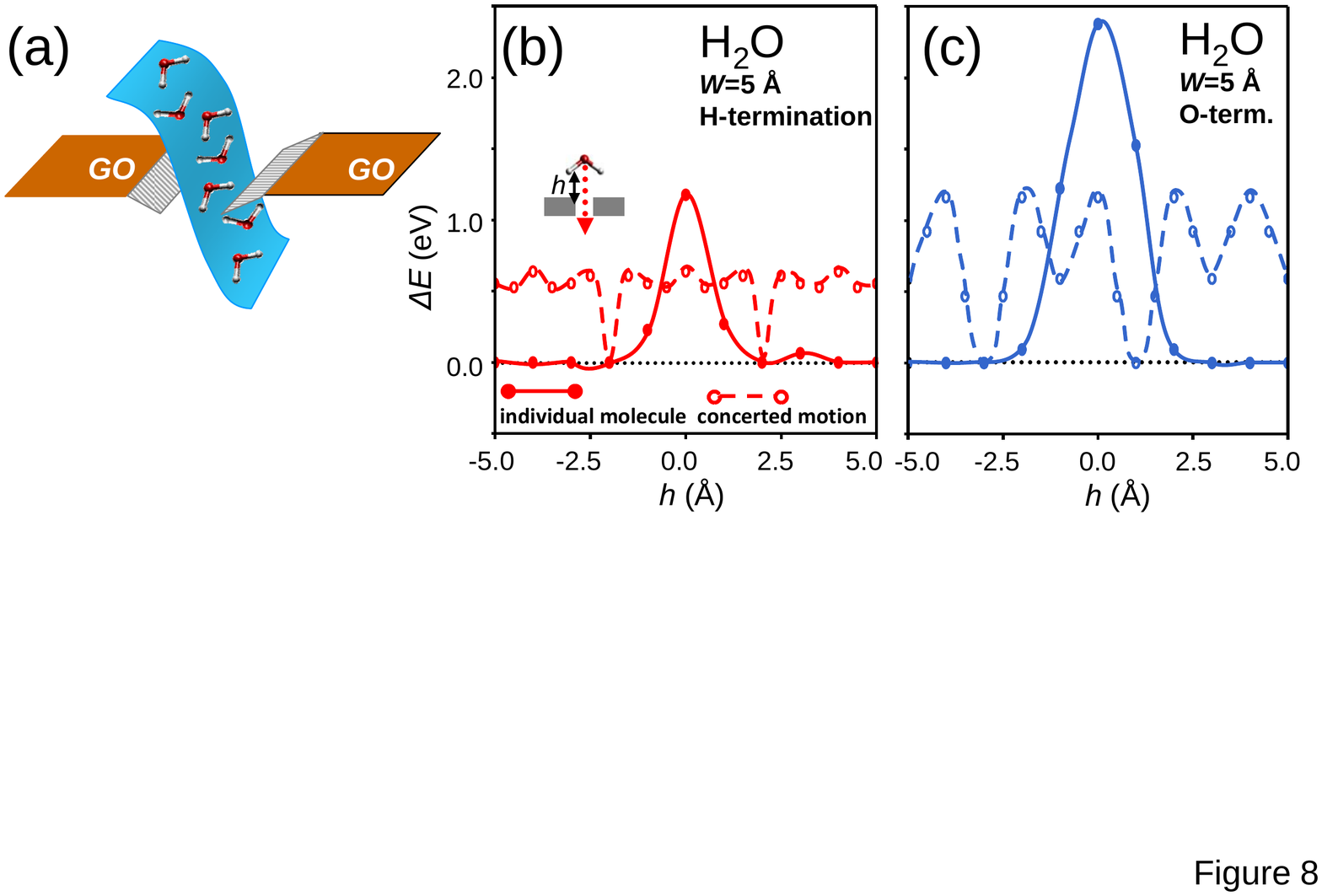}
\caption{%
(a) Schematic of the concerted motion of water molecules across a
chemically terminated in-layer pore in GO. %
Calculated energy change ${\Delta}E$ per H$_2$O permeating either
individually or in concerted manner through
an in-layer pore of width $W=5$~{\AA} with %
(b) hydrogen-terminated and %
(c) oxygen-terminated armchair edges. %
$h$ represents the vertical distance between the centroid of
H$_2$O and the closest carbon layer. Concerted motion of
molecules, indicated by the open symbols and the dashed lines, is
compared to the passage of individual molecules, indicated by the
solid symbols and the solid lines. Lines connecting the simulation
data points are guides to the eye.
} %
\label{fig8}
\end{figure*}

\begin{video}[t]
\includegraphics[width=0.7\columnwidth]%
{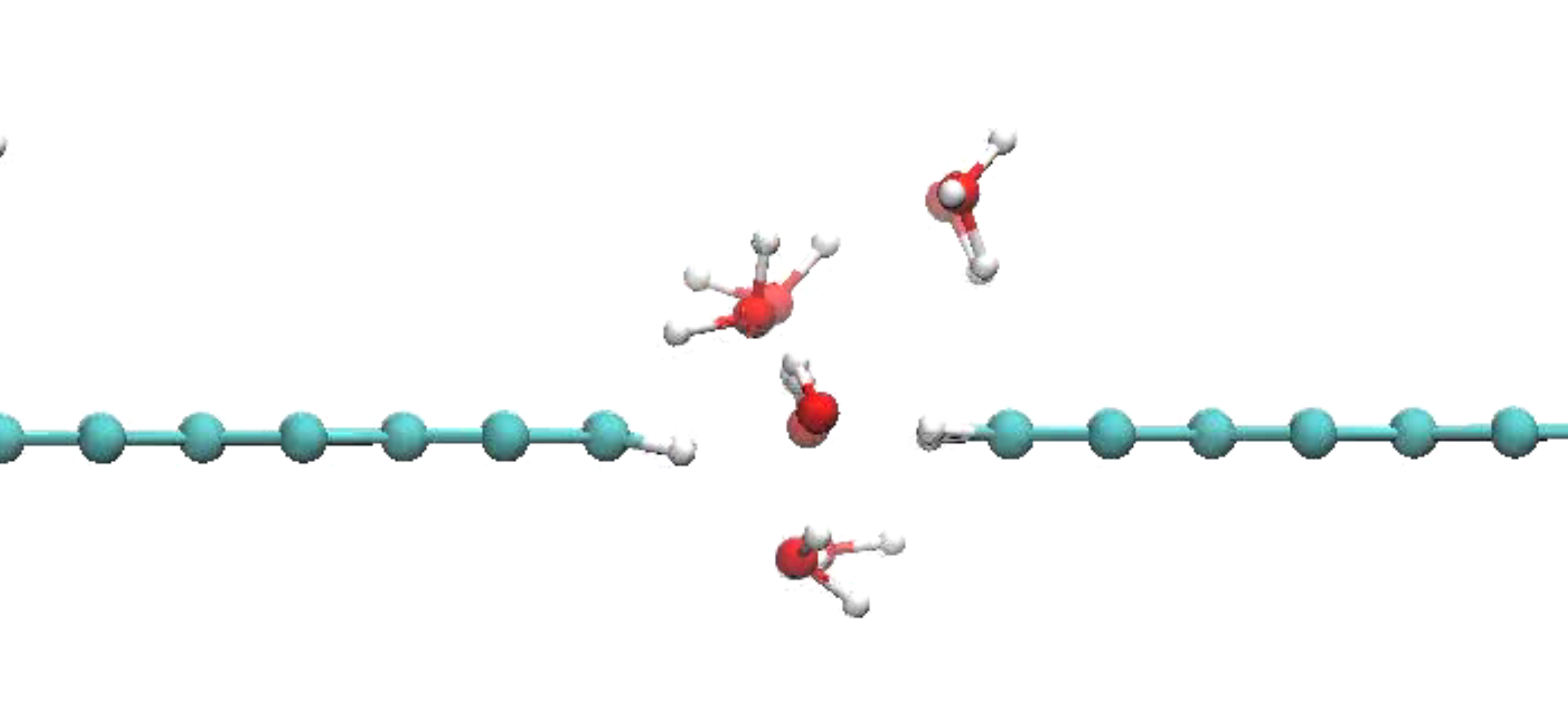}
\setfloatlink{video1.mp4} %
\caption{%
Concerted motion of water molecules permeating across an in-layer
pore, represented by a pair of hydrogen-terminated armchair
graphene nanoribbons separated by $W=5$~{\AA}.
} %
\label{video1}
\end{video}

According to Figs.~\ref{fig7}(c) and \ref{fig7}(d), the activation
barrier for water passage drops down to zero in wider pores with
$W=9$~{\AA}, whereas the activation barrier for hydrated Na$^+$
ions remains substantial with the exception of O-terminated zigzag
edges. As seen in the snapshots from the optimized trajectory in
Figs.~\ref{fig7}(c) and \ref{fig7}(f), Na$^+$ ions lose most of
their hydration shell during the passage through the pore, which
accounts for much of the energy cost. As expected, the energy
difference ${\Delta}E(h)$ is symmetric around $h=0$.

Taking into the account the van der Waals radius of
hydrogen~\cite{Bondi1964} of $1.2$~{\AA}, the effective size of an
in-layer armchair pore, shown in Fig.~\ref{fig6}(a), is reduced
significantly in presence of terminating hydrogen atoms. Thus,
terminating hydrogens reduce the effective opening of a
$W=7$~{\AA} wide pore to $4.6$~{\AA} and that of the even narrower
$W=5$~{\AA} pore to merely $2.6$~{\AA}, less than the van der
Waals diameter $d_{vdW}$(H$_2$O)$=2.8$~{\AA} of a H$_2$O
molecule~\cite{Jorgensen83}. In the latter case, the passage of a
single H$_2$O molecule requires local deformation of the pore at
the energy cost ${\Delta}E=1.1$~eV.
Most favorable deformations at the pore edge will likely involve
bending and not stretching or compression of C-H bonds, as
indicated by observed IR-frequencies in aromatic hydrocarbons,
which suggest that bending a $C_{ar}-H$ bond requires three times
less energy than stretching this bond~\cite{Silverstein2014}. The
same rule applies for other terminating groups as well. Therefore,
in order to allow a water molecule subject to an external force to
pass through an in-layer pore, the terminating atoms at the two
sides of pore will preferably deflect in the same direction, as
indicated in Fig.~\ref{fig8}(a), in much the same way as a saloon
door swings open when a drunk patron is thrown out.

In a process benefiting water, but not solved ions, energetically
activated passage of H$_2$O molecules through narrow in-layer
pores may be further promoted by concerted motion involving many
molecules connected by hydrogen bonds as discussed in the
following Subsection.

\subsection{Concerted motion of H$_2$O molecules through
            in-layer pores in GO}

Understanding the passage of a single water molecule across the
pore is of limited value in view of the fact that water molecules
are connected in a network of hydrogen bonds. A more realistic
scenario considers many water molecules passing through the
in-layer pore at the same time, like a waterfall, depicted
schematically in Fig.~\ref{fig8}(a). Instead of one water
molecule, we have considered a 2D array of molecules confined in a
plane normal to and containing the pore in GO. It is appealing to
consider water molecules assisting each other energetically in the
passage through an in-layer pore by preventing the relaxation of
the terminating edges to the `closed' position, same as saloon
patrons may keep the swing door open for each other when entering.
Our results, shown in Fig.~\ref{fig8}(b) for a hydrogen-terminated
and in Fig.~\ref{fig8}(c) for an oxygen-terminated edge, indicate
that concerted motion of water molecules may indeed reduce the
energy barrier for their passage from $1.1$~eV to $0.6$~eV %
in case of hydrogen termination and from $2.3$~eV to $1.3$~eV for
oxygen termination, almost to half the original value. Movies of
the concerted motion of H$_2$O molecules across in-layer pore are
shown in Video~\ref{video1}
and in Section C of the SI.


\subsection{Permeation of different ions through
            in-layer pores in GO}

\begin{figure}[t]
\centering
\includegraphics[width=1.0\columnwidth]{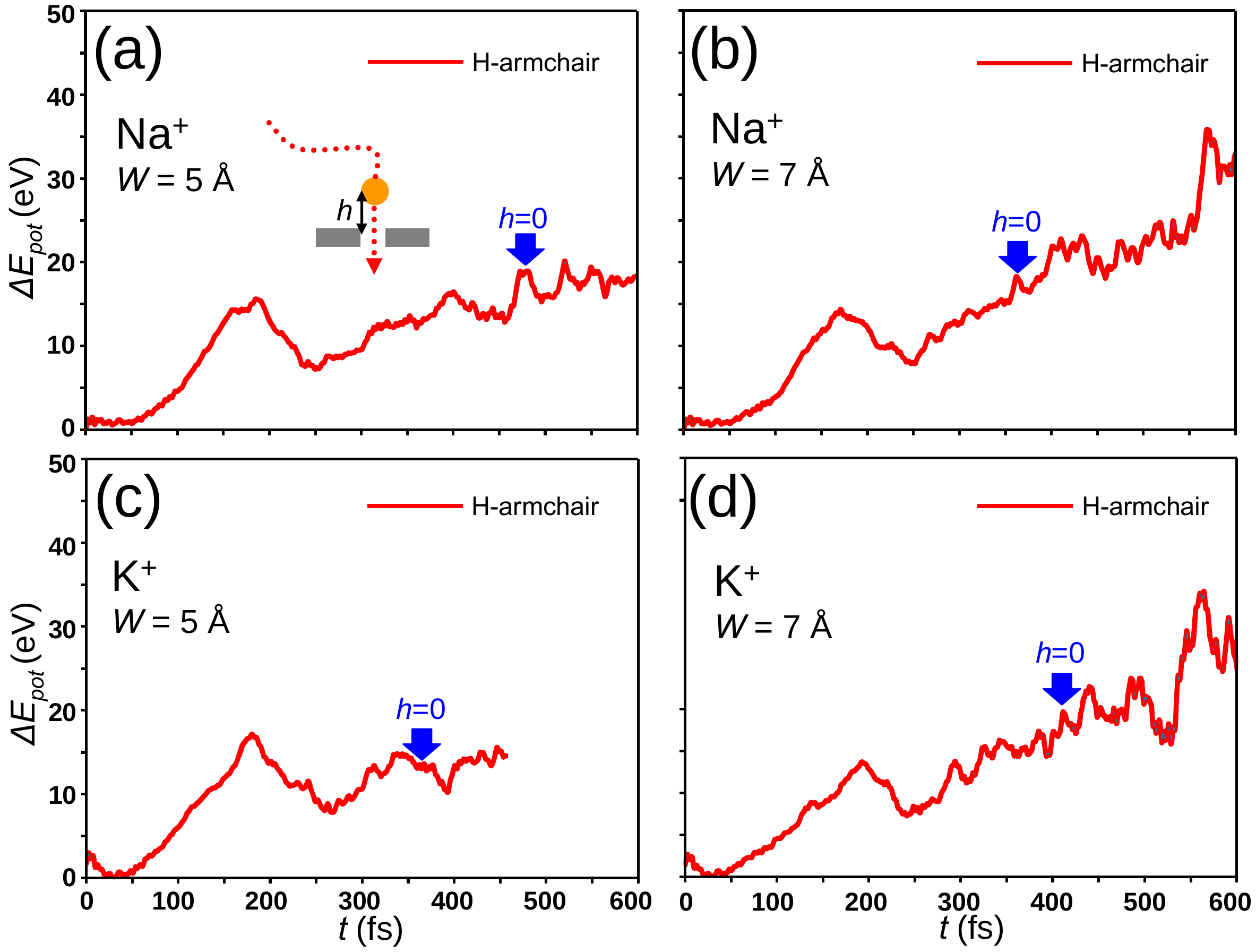}
\caption{%
Changes in the potential energy of the system ${\Delta}E_{pot}$ as
fully hydrated Na$^+$ (a,b) and K$^+$ (c,d) ions are driven by an
applied force to pass through a hydrogen-terminated in-layer pore
in graphene with armchair edges. Simulation results for pores of
width $W=5$~{\AA} in (a,c) are presented next to results for wider
pores with $W=7$~{\AA} (b,d). Each unit cell contains one ion and
28 H$_2$O molecules. $h$ represents the vertical distance between
the ion and the closest carbon layer.
} %
\label{fig9}
\end{figure}

As mentioned above, both isolated and hydrated K$^+$ ions are
significantly larger than Na$^+$ ions. Nevertheless, K$^+$ ions
are known to be transported across protein membranes in preference
to Na$^+$ ions~\cite{{Doyle1998},{Zhou2001}}. Even though protein
membranes are significantly more complex that in-layer pores in
GO, the postulated reason reason for this behavior being the
stronger hydration shell of Na$^+$ as compared to K$^+$
ions~\cite{Sadhu2015} applies in both cases.

We compared permeation of Na$^+$ and K$^+$ ions in water across
hydrogen-terminated in-layer armchair pores in graphene with
$W=5$~{\AA} and $W=7$~{\AA}. To simulate this process, we
performed microcanonical MD simulations of the system that had
been initially equilibrated at $T=300$~K. We monitored changes in
the potential energy ${\Delta}E_{pot}(t)$ of the system as a
function of time. During the run, all water molecules and ions
have been subject to an external force characterized by
$|{\bf{F_0}}|=6.25{\times}10^{-3}$~eV/{\AA},
applied normal to the graphene layers. No constraints have been
applied during the $0.6$~ps-long simulations except for fixing all
carbon atoms in their positions. Results of our single-trajectory
simulations are presented in Fig.~\ref{fig9}. We should note that
these results should be considered only illustrative of the
process and that ensemble averages over many trajectories would be
required for any quantitative conclusions.

As expected, the action of the external force gradually increases
the total energy and, in view of the virial theorem, also the
potential energy of the system.
Geometries corresponding to relative minima in
${\Delta}E_{pot}(t)$ graphs are to be considered more stable. We
found it interesting that the point of ion penetration across the
layer, defined by $h=0$, is not prominent in any of the graphs.
This indicates that the relative position of water molecules with
respect to the pore plays as important role as that of the ions.
In view of the uncertainty caused by the finite size and complex
structure of the system, ions pass faster through wider pores than
through narrower pores. We found relative maxima in
${\Delta}E_{pot}(t)$ in all trajectories at $t{\lesssim}200$~fs
from the beginning and associate these peaks with the necessity of
structural rearrangements in water molecules adjacent to the ions
and the pore that would subsequently enable the passes of the ion
across the pore. With respect to this configuration prior to
passage through the pore, the one corresponding to $h=0$ may be
either more stable as in Fig.~\ref{fig9}(c), or less stable, as in
the other panels. Based on these preliminary results, we can not
find any conclusive evidence for preferential passage of Na$^+$ or
K$^+$ ions through in-layer pores in graphene or GO.


\section{Discussion}

The major objective of our study was to find whether GO is a
viable material for water desalination~\cite{Boehm1961} as
postulated by many since the 1960's. To mitigate the brittle
nature and related shortcomings of GO, we postulated a sandwich
structure containing ultra-strong carbon nanotubes and graphitic
carbon fabric in the outer layers enclosing the GO core layer. One
of the key motivations for our study has been the urgent call for
a paradigm shift in the desalination membrane community, largely
unaware of recent progress in ultra-strong carbon nanotubes, and
the desire of the carbon nanotube community to locate a `killer
application'. Based on what has been known about each component of
the system, and based on our microscopic {\em ab initio}
calculations detailing atomic-scale processes occurring in the GO
layer, we find the proposed all-carbon membrane well worth
considering seriously for water desalination and encourage its
experimental realization.

The proposed membrane design addresses the most urgent problems in
water desalination, namely selective salt rejection falling short
of its objective~\cite{Werber2016a} and irreversible bio-fouling.
We have identified structural parameters that should maximize salt
rejection by GO without sacrificing water permeability. We believe
that mechanical treatment of GO by ball milling and shear
alignment may provide a uniform system well suited for its
purpose. Graphitic carbon nanostructures forming the layers
surrounding GO are now well established for their mechanical
strength, flexibility, electrical and thermal conductivity, and
resilience to extreme heat and to chemicals. Very important is the
fact that all components of the all-carbon membrane are
commercially available at very low cost.

Selective rejection of ions and reduction of its bio-fouling
propensity may result from charging the conductive membrane
possibly by a pulsed current~\cite{Ho2018}. Additional resistance
against bio-fouling may be provided by continuously changing the
morphology of the outermost layers of the proposed all-carbon
membrane~\cite{Pocivavsek18}. This should be possible, because
individual components are graphitic structures that, under
compression, may kink or wrinkle, but not uniformly shrink due to
the high in-plane rigidity or graphene. Since reverse osmosis
membranes are not static objects, but rather vibrate under
operating conditions, the membrane surface will continuously
change its morphology and thus detach debris and bio-fouling
agents~\cite{Pocivavsek18}.

Key to selective rejection of ions and permeation of water
molecules is the morphology and chemical treatment of the central
GO layer. The hydrophilic nature of GO can be regulated by
modifying the density of OH terminating groups. The size and
nature of in-layer pores, which we find critical for the
desalination process, can initially be optimized during the ball
milling and shear alignment process. Membranes with fine pores are
suitable for deionizing water, whereas coarse-pored membranes will
find their use in multi-stage RO systems or in decreasing the
salinity of brines. Besides their smaller size compared to
hydrated ions, water molecules benefit from their ability to pass
through even very narrow pores by concerted motion with quite low
activation barriers. Even $5$~{\AA} wide pores are still permeable
while providing high salt rejection.

According to our results, $7$~{\AA} pores provide the best
permeability ratio between H$_2$O an Na$^+$, whereas still wider
$9$~{\AA} pores provide lower selectivity. The mechanical strength
and flexibility of graphitic carbon allows adjusting the operating
pressure, which should modify the pore size in order to adjust
salt rejection and water flux as needed. Measuring and feeding
these data continuously to a control system should allow
on-the-fly tuning of operating pressure in response to changing
feed water parameters~\cite{Lior2012p494}. Pores should open at
higher and close at lower pressures, modifying the performance of
the membrane. The range of operating parameters, in particular
pressure, which optimize specific aspects of the membrane
performance, can be studied in pilot plants and provided to the
consumer by the manufacturer.

According to our simulations and common chemical intuition,
presence of OH groups is largely responsible for turning
hydrophobic graphite into hydrophilic GO. As shown in our study,
hydrophilic interlayer slit pores in GO attract ions and reduce
their mobility, thus further decreasing permeation of salt in a
similar way as in-layer pores do~\cite{Yang16}. For this reason,
increasing the number of OH groups at the expense of epoxy-groups
is highly desirable from the point of view of salt rejection.

As mentioned in the text, edges of in-layer pores are unlikely to
remain bare and will acquire terminating groups under ambient
conditions. We studied three likely candidates for chemical
termination, namely H, epoxy-O and OH groups, but did not observe
significant differences from the viewpoint of water and salt
permeation except a minor preference for hydrogen termination.
Given a fixed in-layer pore width $W$, hydrogen termination
hinders entering water molecules least due to its small size.
Also, the activation barriers for Na$^+$ passage through
H-terminated pores are not the highest. However,
hydrogen-terminated pores provide the highest
${\Delta}E$(H$_2$O)/${\Delta}E$(Na$^+$) ratio, which means they
maximize the permeability difference -- the quantity we strive to
increase. Hydrogen termination is also favored, because C-H groups
interact only weakly with dissolved ions.

As we can infer from Fig.~\ref{fig7}, Na$^+$ ions appear to be
stabilized inside some O- and OH-terminated pores, resulting in a
``negative barrier height''. The reason for this behavior is the
arrangement of O atoms in these pores, which provides an excellent
coordination sphere for sodium cations, resembling the structure
the ion channels selectivity filter~\cite{Doyle1998}. If the
cation binding is too strong, the first cation that enters the
pore would clog it. If the binding is weaker, small energetic
preference for the in-pore site may promote cation passage by the
adsorption-desorption mechanism. Unlike O- and OH- terminating
groups, terminating hydrogens act mostly like a mechanical brush
without the potential negative side effects.


\section{Summary and Conclusions}

We have designed an all-carbon membrane for the purpose of
filtering and desalinating water. The membrane contains graphite
oxide (GO) sandwiched in-between layers of buckypaper consisting
of carbon nanotubes, and the entire structure is contained
in-between layers of strong carbon fabric from both sides. The
structure combines high mechanical strength with thermal
stability, resilience to harsh chemical cleaning agents with
electrical conductivity, thus addressing major shortcomings of
commercial reverse osmosis membranes. We have used {\em ab initio}
DFT calculations to provide atomic-level insight into the
permeation of water molecules in-between GO layers and across
in-layer vacancy defects. Our calculations elucidate optimum
conditions for permeation of water and selective rejection of
solvated Na$^+$ ions by the membrane. We anticipate that the
viability of the designed structure will be confirmed
experimentally in the near future.


{\bf Note added in proof.} We have recently learned of a different
all-carbon desalination membrane, which has demonstrated superior
performance over current technology~\cite{Yang19}.


\begin{acknowledgments}
D.T. acknowledges financial support by the NSF/AFOSR EFRI 2-DARE
grant number EFMA-1433459. A.K. acknowledges financial support by
the Fulbright program. We thank Igor Baburin, Morinobu Endo,
Katsumi Kaneko, Takeyuki Kawaguchi, James Linnemann, Dan Liu,
Alexander Quandt, Gotthard Seifert, Alexandr Talyzin, Akihiko
Tanioka and Volodymyr Tarabara for valuable discussions.
Computational resources have been provided by the Michigan State
University High Performance Computing Center.
\end{acknowledgments}


%


%

\end{document}